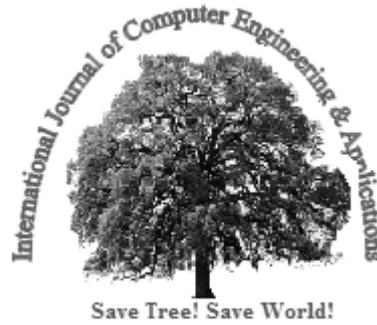

# PREDICTION OF RATE OF IMPROVEMENT OF SOFTWARE QUALITY AND DEVELOPMENT EFFORT ON THE BASIS OF DEGREE OF EXCELLENCE WITH RESPECT TO NUMBER OF LINES OF CODE

### Ekbal Rashid, Srikanta Patnaik, Vandana Bhattacherjee

*Department of CS & E, C.I.T , Tatisilwai, Ranchi*
*Department of CS & E, SOA University, Bhubaneshwar, Orissa*
*Department of CS & E, B.I.T, Mesra, Ranchi*

**ABSTRACT:**

*The objective of this research work is to improve the degree of excellence by removing the number of exceptions from the software. The modern age is more concerned with the quality of software. Extensive research is being carried out in this direction. The rate of improvement of quality of software largely depends on the development time. This development time is chiefly calculated in clock hours. However development time does not reflect the effort put in by the developer. A better parameter can be the rate of improvement of quality level or the rate of improvement of the degree of excellence with respect to time. Now this parameter needs the prediction of error level and degree of excellence at a particular stage of development of the software. This paper explores an attempt to develop a system to predict rate of improvement of the software quality at a particular point of time with respect to the number of lines of code present in the software. Having calculated the error level (EL) and degree of excellence (DE) at two points in time, we can move forward towards the estimation of the rate of improvement of the software quality with respect to time. This parameter can estimate the effort put in while development of the software and can add a new dimension to the understanding of software quality in software engineering domain. In order to obtain the results we have used an indigenous tool for software quality prediction and for graphical representation of data, we have used Microsoft office 2007 graphical chart.*

## [1] INTRODUCTION

The main thrust in modern software engineering research is centered on trying to build tools that can enhance software quality. Software quality estimation models in software engineering are used to predict important attributes such as software development effort, software reliability, and productivity of programmers [2]. Software quality prediction is a complex mix of characteristics and varies from application to application and users who requests for it. A software quality prediction model can be used to identify the defective modules [3]. Although cost estimation and quality estimation may have relative independence, the two are dialectically dependent on one another. Cost reduction can be to some extent considered as a parameter of quality. At the same time, quality improvement is sure to affect the cost factor. The cost of software is related to the development time. However the development time does not truly reflect the effort put in by the developer. A better method is to calculate the rate of improvement of software quality and the effort put in by a developer and to use it as a new parameter to provide a better understanding of the cost of a software. This paper focuses upon the need and the methods to estimate the rate of improvement of software quality and the effort put in by the developer in the course of software development. Rashid et. al emphasized on the importance of case-based estimation model for software quality prediction [4].

The rest of the paper is structured as follows: section II gives the problem description, section III describes the significance of the study, section IV describes the brief overview of error, section V describes the methodology, section VI describes why testing is important. In section VII we present the analysis and result. A conclusion has been presented in section VIII.

## [2] PROBLEM DESCRIPTIONS

The main problem in trying to calculate the rate of improvement of software quality and the effort put in by a developer cannot be measured in a straightforward manner like measuring clocked development time or counting the lines of code. The concept rate of improvement of quality or of effort of development seems to be too abstract as compared to the above mentioned parameters. It has to be understood and defined unambiguously before proceeding further. There may be many arguments and views on the question as to what the term effort actually means and also on what should be the method of its calculations. Nevertheless, a beginning is necessary for refinements to follow. In due course of time, the best understanding may evolve as a result of the collective efforts of many. The problem of estimating the rate of improvement of quality and in due course calculating the effort as a part of trying to evaluate the cost and the quality of a software makes it necessary to record the history of software development and that too with objective precision. Naturally the question arises as to what is to be recorded? One can record the number of lines of code at a particular instance of time. But this again may not reflect a correct picture of development status. Suppose there are thousands of lines of code with thousands of errors. On the other hand there may be a few hundred lines of error free code. Quite understandably the thousands of lines of code with as many errors will surely have a much more negative impact on the software development than the few hundreds of lines of error free code.

Hence the question of quality of software is somewhat related to the rate of improvement at a particular instance of time. That needs to be correctly ascertained.

The main focus of the paper is thus the estimation of the rate of improvement of software quality.

## [3] SIGNIFICANCE OF STUDY

The method of calculation of the rate of improvement of quality and in due course the effort estimation as a parameter to calculate the value of the software may develop newer ideas and understanding in the field of software engineering. Treatment of an abstract entity as effort in development may transform to many advanced concepts of what should be the value created by the developer. This may on the one hand decide the developer's merits and strengths while on the other hand may provide a guideline to set a suitable price for the developed software. Although many other factors are inexorably linked in deciding a suitable price for the software, yet the effort parameter shall play no less than a significant role in this regard.

There is another significance that goes with this study. The calculation of error level and the degree of excellence as an objective measure of the software quality at a particular instance of time is a novel application, unused till to date, or at least to their knowledge used very insignificantly. This means an applied form has been successfully evolved from a theoretical construct to enhance the implementation of software engineering concepts.

Again the measure of error level and the degree of excellence calls for the deployment of an indigenous tool that can calculate the required statistics accurately and precisely. This again opens up newer vistas in the already vast areas of compiler design and the integration of compiler design into the engineering of automated software. Notwithstanding the indigenous tool again explores further possibilities in this direction
.

## [4] ERRORS

Error refers to a condition that creeps in unwanted. It is a mistake committed by a person due to factors which are detrimental to his or her value. It may be lack of concentration, lack of proper training, improper conception of the issue concerned, anxiety, stress and many other factors. Whatever may be the reason, human error is bound to be associated with any piece of work and the conception debugging has actually generated from the understanding that a scientific understanding has to be developed to remove the errors. Now the number of errors that a less experienced person can make is likely to be more that the number of errors likely to be made by a more experienced hand. This brings in the necessity to minimize the number of errors found in any particular piece of work.

## [5] METHODOLOGY

Before going into the methodology, here are some basic definitions and principles [1]:
1.      Error Level (EL):
The error level of software at a particular instance of time with respect to the number of lines of code (LOC) is defined as the ratio of the number of errors detected to the number of lines of code at that particular instance of time.

$$EL = \frac{number\ of\ errors\ detected}{number\ of\ LOC}$$

Now this ratio when expressed in percentage gives the percentage error level.
2.      Degree of Excellence (X):
The Degree of Excellence of software at a particular instance of time with respect to the number of lines of code is defined as the ratio of the correct number of lines of code to the total number of lines of code expressed in percentage.
Thus:

$$X = \frac{\text{correct number of LOC}}{\text{total number of LOC}}$$

This is also equivalent to:-

X = 100 – EL%

3.      Improvement of degree of Excellence:

The improvement of the Degree of Excellence of software may be defined as the difference between the Degrees of Excellence between two points in time.

Thus:

Improvement = $X_f$ - $X_i$

Where $X_f$ and $X_i$ stand for final Degree of Excellence and initial degree of Excellence between two points of time.

4.      Rate of improvement of software quality:

It can be easily understood that once the developer or the team of developers has been decided upon the parameter X (Degree of Excellence) is just a function of time. And then we can define the rate of improvement of software quality as:

The rate of improvement of software quality is the derivative of the degree of excellence with respect to time. With this understanding the following principle can be stated:

5.      Principle of Effort in Software Development:

The principle of effort in software development states that –

'The effort put in software development is proportional to the rate of improvement of degree of excellence of the software with respect to time. '

Thus:

$$E \propto \frac{dX}{dt}$$

or

$$E = \alpha \frac{dX}{dt}$$

where α is the constant of proportionality and may be called the coefficient of the developer's ability.

The methodology involved is the following:

Using an indigenous tool, we first calculate the error level and degree of excellence at two different points of time. Then the improvement of the degree of excellence and subsequently the rate of improvement of the degree of excellence can be calculated in the methods mentioned above. If the improvement of the degree of excellence is expressed as a function of time, then the effort in software development at a particular instance of time can be calculated as the derivative of the degree of excellence with respect to time at that particular point of time. The indigenous tool uses 'C' as the programming language.

## [6] WHY TESTING IS IMPORTANT

Due to the activities connected with the fault detection account for a considerable part of the project budget. Examining a software module and predicting whether it is faulty or non faulty. A faulty module is a module which has not passed the certain number of test cases. We have developed a model for software testing (see figure 1).

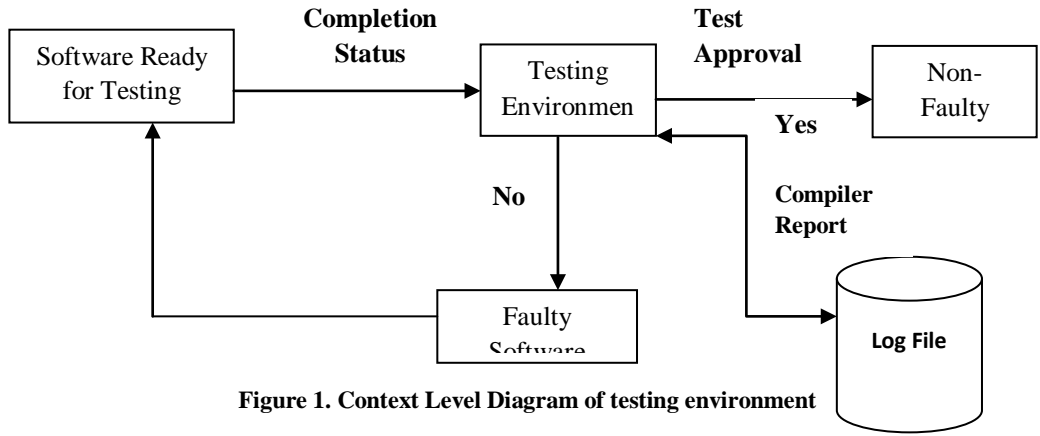

**Figure 1. Context Level Diagram of testing environment**

# [7] ANALYSIS AND RESULTS

The aim of this research was to identify the number of errors in the software as well as total number of loops used in the software like for loop, while loop. The objective was to test the software whether the software is faulty or non faulty on the basis of results given by the indigenous tool.

The tool was working as:

1) The tool asks for the file which contains the code and after that has been provided,

2) The tool scans for the number of lines of code (LOC) excluding comments line in the software after that is done,

3) The tool asks for the name of the log file which contains the error report of the compiler.

4) After the name of the log file has been provided, the tool searches for the number of errors in the log file and then calculates the degree of excellence as well as the error level of the code at the particular instance of time and gives the output (see Screenshot1 through Screenshot 2).

```
C:\WINDOWS\system32\cmd.exe - newsoft2

Enter the name of the file : four.c
File opened successfully!
The number of lines in the file is : 675
Number of comment lines is : 67
The number of for loops is : 20
The number of while loops is : 4
Number of errors = 0
loc = 608
Error level w.r.t LOC = 0.00
Quality Level or Degree of excellence = 100.00

Want to continue? y/n : _
```

**Screenshot 1: Quality Level or Degree of excellence of a file with no error is 100 percent**

Ekbal Rashid, Srikanta Patnaik and Vandana Bhattacherjee

```
C:\WINDOWS\system32\cmd.exe - newsoft2

Enter the name of the file : exponent.c
File opened successfully!
The number of lines in the file is : 1117
Number of comment lines is : 173
The number of for loops is : 27
The number of while loops is : 4
Number of errors = 8
loc = 944
Error level w.r.t LOC = 0.85
Quality Level or Degree of excellence = 99.15

Want to continue? y/n :
```

Screenshot 2: analysis of file with eight errors, Quality Level or Degree of excellence 99.15 percent

The output can be noted down and a graph can be prepared as shown below. The different types of graph forms show uniform or constant improvement, positive improvement and negative improvement respectively. The slope of the secant of the graph at any two points of time gives the average improvement between those two points in time while the tangent at any point gives the instantaneous rate of improvement at any particular instance of time shown in fig. 2.

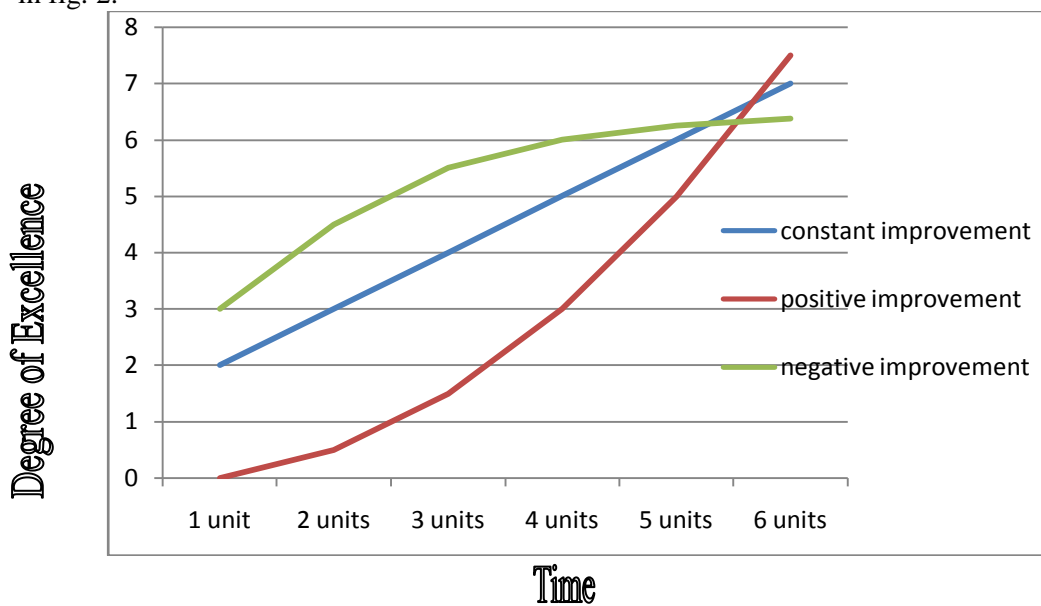

Figure. 2. Degree of Excellence with respect to time

## [8] CONCLUSION

The calculation of the rate of improvement of quality and effort in the software development may add new dimensions to the understanding of software engineering. The paper will be expanded to include newer concepts on the basis of the understanding of the rate of

11

improvement of quality and Effort in software development which no longer remains an abstract term but on the contrary becomes a true quantity that can be measured, estimated and compared at different levels. The knowledge about the rate of improvement of software quality and Effort in software development, leads us to the correct estimation of the quality of the software and also the quality of the developer. The price of the software can be ascertained more precisely as a result. The future target in this regard is the development of fully automated software that accepts the name of the file which contains the code of the software and calculates the Degree of Excellence at a particular point of time. Then the software will calculate the average rate of improvement of quality and Effort and using suitable methods also formulate the same as a function of time. Once these quantities have been expressed as a function of time, the derivative and as a result the instantaneous Effort can be calculated. Further research also aims towards dealing exhaustively on the coefficient of ability of the developer as mentioned in the present paper. It will explore the fields on which this coefficient of ability depends and define it in a much more concrete manner. Effort no longer will remain a subjective entity. It will assume objective features.

**REFERENCES**

[1]  Mordechai Ben-Menachem, Garry S. Marliss "Software Quality Producing Practical, Consistent Software, Thomson Learning.

[2]  V. Bhattacherjee and S. Kumar,(2004),"Software cost estimation and its relevance in   the Indian software Industry", In Proceedings of the International Conference on Emerging Technologies IT Industry, Nov'05, PCTE, Ludhiana India.

[3]  Ekbal Rashid, Srikanta Patnaik, Vandana Bhattacherjee (2012)"A Survey in the Area of Machine Learning and Its Application for Software Quality Prediction" has been published in ACM SigSoft ISSN 0163-5948, volume 37, number 5, September 2012, http://doi.acm.org/10.1145/2347696.2347709 New York, NY, USA.

[4]  Ekbal Rashid, Srikanta Patnaik, Vandana Bhattacherjee "Enhancing the accuracy of case-based estimation model through Early Prediction of Error Patterns" proceedings published by the IEEE Computer Society 10662 Los Vaqueros Circle Los Alamitos, CA, in International Symposium on Computational and Business Intelligence (ISCBI 2013), New Delhi, 24~26 Aug 2013 ISBN 978-07695-5066-4/13 IEEE, pp. 208-212, DOI 10.1109/ISCBI.2013. http://ieeexplore.ieee.org/xpl/tocresult.jsp?isnumber=6724301.